\documentclass[amsfonts,amssymb,twocolumn,showpacs,nofootinbib]{revtex4-1}

\usepackage{amsmath}
\usepackage{amsfonts}
\usepackage{amssymb}
\usepackage{mathrsfs}
\usepackage{amsthm}
\usepackage{wasysym}
\usepackage[colorlinks]{hyperref}
\usepackage[all]{hypcap}
\usepackage{graphicx}
\usepackage[usenames]{color}
\usepackage{xspace}
\usepackage{verbatim}
\usepackage{enumerate}

\usepackage[T1]{fontenc}
\usepackage{microtype}

\newcommand{\nn}{\nonumber}

\newcommand{\pd}{\partial}

\newcommand{\pont}{{}^{*}\!RR}

\DeclareSymbolFont{euletters}{U}{eur}{m}{n}
\DeclareMathSymbol{\wp}{\mathord}{euletters}{"7D}

\begin{document}

  \title{Note on Legendre decomposition of the Pontryagin density in Kerr}
  \author{Leo C. Stein}
  \thanks{Einstein fellow}
  \email{leostein@astro.cornell.edu}
  \affiliation{Center for Radiophysics and Space Research, Cornell
    University, Ithaca, NY 14853 USA}

  \date{\today}

  \begin{abstract}
    In~\cite{Konno:2014qua} (``Scalar field excited around a rapidly
    rotating black hole in Chern-Simons modified gravity''), Konno and
    Takahashi have recently developed some analytical results for the
    scalar field about a Kerr black hole in the decoupling limit of
    dynamical Chern-Simons gravity. This involved a decomposition of
    the source (the Pontryagin density) in terms of Legendre
    polynomials. Here we give a two-line expression for this
    decomposition which simplifies their quadruple sum. Our
    expressions are rational polynomials multiplying Legendre
    functions of the second kind, or equivalently rational polynomials
    multiplying hypergeometric functions.
  \end{abstract}


 \maketitle

In~\cite{Konno:2014qua}, Konno and Takahashi (henceforth KT) attempted to find
an analytical solution for the scalar field $\vartheta$ which
satisfies
\begin{equation}
  \label{eq:box-theta}
  \square\vartheta \propto \pont
\end{equation}
where $\pont=-{}^{*}\! R^{abcd} R_{abcd}= -\frac{1}{2}\epsilon^{abef}
R_{ef}{}^{cd} R_{abcd}$ is the Pontryagin density. This is one of the
equations of motion of dynamical Chern-Simons (dCS)
gravity~\cite{Jackiw:2003pm,Alexander:2009tp}, the other being the
modified Einstein equation, which also includes corrections from
$\vartheta$. In the decoupling limit about some background (GR)
solution, the wave operator of Eq.~\eqref{eq:box-theta} becomes the
background wave operator and the source term is evaluated on the
background spacetime. When seeking rapidly-rotating black hole
solutions we use the Kerr spacetime~\cite{Kerr:1963ud,Visser:2007fj}
as the background solution, and impose stationarity and axisymmetry on
$\vartheta$. Then we desire to solve
\begin{equation}
\label{eq:stationary-axisymmetric-Sigma-box}
  \left[
\frac{\pd}{\pd r} \left(\Delta \frac{\pd}{\pd r}\right)
+  \frac{1}{\sin\theta}\frac{\pd}{\pd\theta} \left( \sin\theta\frac{\pd}{\pd\theta} \right)
  \right]\vartheta = C \Sigma \pont \,,
\end{equation}
where $\Delta\equiv r^{2}+a^{2}-2Mr$, $\Sigma\equiv
r^{2}+a^{2}\cos^{2}\theta$, $C$ is some constant related to the
coupling strength of dCS, and in Kerr we have
\begin{equation}
\label{eq:pont}
  \pont = \frac{96 M^{2}a c r}{\Sigma^{6}}
  (3r^{2}-a^{2}c^{2})(r^{2}-3a^{2}c^{2})\,,
\end{equation}
where $c=\cos\theta$.
\newpage
The differential operator on the LHS of
Eq.~\eqref{eq:stationary-axisymmetric-Sigma-box} separates, and the
eigenfunctions of the homogeneous angular equation are simply the
Legendre polynomials $P_{l}(c)$.
Therefore we consider the decomposition
\begin{align}
  S &\equiv C \Sigma\pont = \sum_{l=0}^{\infty} S_{l}(r)
  P_{l}(\cos\theta)
\end{align}
where
\begin{align}
  \label{eq:source-decomp-def}
  S_{l}(r) &= \frac{2l+1}{2} \int_{-1}^{+1} S(r,c) P_{l}(c) dc \,.
\end{align}
The source is odd in $c$, so we will only need to consider odd $l$.

For simplicity we define
\begin{align}
  \label{eq:source-decomp-def-explicit}
  S_{l}(r) &= 96 C M^{2} \frac{2l+1}{2} I_{l}(r) \\
  \label{eq:I-l-def}
  I_{l}(r) &\equiv
  \int_{-1}^{+1}
  \frac{a c r (3r^{2}-a^{2}c^{2}) (r^{2}-3a^{2}c^{2})}{(r^{2}+a^{2}c^{2})^{5}}
  P_{l}(c) dc\,,
\end{align}
which appears as a source term for the separated inhomogeneous radial
equation for $\vartheta_{l}(r)$, defined similarly to $S_{l}(r)$.

\section{Two-line expression for $I_{l}(r)$}
\label{sec:integrating}
KT give their $S_{2n+1}(\tilde{r})$ (not quite equal to ours, see
Sec.~\ref{sec:checks-comparisons}) in their Eqs.~(29-36) in terms of a
quadruple sum of rational polynomials and the $\arctan$ function.  We
find two-line expressions for $I_{l}(r)$ in terms of special
functions:
\newcommand{\irona}{{\textstyle\frac{ir}{a}}}
\begin{widetext}
  \begin{align}
\label{eq:I_l-using-Q}
I_{l}(r) ={}&
-\frac{(l+1)}{24(a^{2}+r^{2})^{4}} \left\{\frac{r}{a}
  \left[a^4 (l^{3}-l^{2}-16l+4)
    +2 a^2 r^{2} \left(l^2 (l+3)-16\right)
    + r^{4} (l+2)^2 (l+3)\right]
  \left[Q_l\left(+\irona \right)+Q_l\left(-\irona\right)
  \right]\right.\nn\\
&\qquad\left.{}+2 \left[a^4
   \left(l^2+l-2\right)-2 a^2 r^{2} \left(l^2+l-8\right)-3 r^{4} \left(l^2+l+2\right)\right]
\frac{1}{i} \left[Q_{l+1}\left(+\irona\right)-Q_{l+1}\left(-\irona\right)\right]\right\}\,,
  \end{align}
or equivalently,
  \begin{align}
\label{eq:I_l-using-F}
I_{l}(r) ={}&
\frac{ \varepsilon(l)\sqrt{\pi } \Gamma (l+2) \, (a/r)^{l}}
{3\, (2)^{l+3}   \left(a^2+r^2\right)^4 \Gamma (l+\frac{3}{2})}
\left\{
  \left[a^4 (l^{3}-l^{2}-16l+4)
    +2 a^2 r^{2} \left(l^2 (l+3)-16\right)
    + r^{4} (l+2)^2 (l+3)\right]
F(\textstyle\frac{l+1}{2},\frac{l+2}{2};l+\frac{3}{2};-\frac{a^{2}}{r^{2}})
\right. \nn\\
&\left.
-\frac{2(l+1)a^{2}}{(2l+3)r^{2}}
\left[a^4 \left(l^2+l-2\right)-2 a^2 r^{2} \left(l^2+l-8\right)-3
  r^{4} \left(l^2+l+2\right)\right]
F(\textstyle\frac{l+2}{2},\frac{l+3}{2};l+\frac{5}{2};-\frac{a^{2}}{r^{2}})
\right\}
\end{align}
\end{widetext}
where $\varepsilon(l)=+1$ for $l=1,5,9,\ldots$, $\varepsilon(l)=-1$
for $l=3,7,11,\ldots$, and $\varepsilon(l)=0$ for even $l$. Below we
describe how to find these expressions.

We make use of the identity%
\footnote{This identity is correct in the fourth
  edition~\cite{1965tisp.book.....G} but incorrect in the seventh
  edition~\cite{2007tisp.book.....G}. I did not have access to other
  editions to check where the error was made.}
\begin{equation}
  \begin{split}
\int_{-1}^{+1} c^k (z-c)^{-1} (1-c^2)^{m/2} P_l^m(c) dc \\
= (+2) (z^2-1)^{m/2} Q_l^m(z) z^k
  \end{split}
\end{equation}
where $m\le l, k=0,1,\ldots l-m$, and $z$ is in the complex plane with
a cut along $(-1,+1)$ on the real axis.
To use this identity, we will have $m=0=k$,
\begin{equation}
\label{eq:Pl-identity}
\int_{-1}^{+1} \frac{P_l(c)}{z-c} dc =  +2 Q_l(z)
\end{equation}
with the same restriction on $z$ as above. To get
Eq.~\eqref{eq:I-l-def} into a form where Eq.~\eqref{eq:Pl-identity}
may be applied, use a complex partial fractions decomposition for the
rational polynomial (i.e.~the denominator $(r^{2}+a^{2}c^{2})^{5}$ is
an irreducible polynomial over $\mathbb{R}$ but it is reducible over
$\mathbb{C}$). The decomposition is
\begin{multline}
  \frac{a c r (3r^{2}-a^{2}c^{2})
    (r^{2}-3a^{2}c^{2})}{(r^{2}+a^{2}c^{2})^{5}} =
\frac{r}{2 (a c - i r)^5} + \frac{r}{2 (a c + i r)^5} \\
-  \frac{i}{4 (a c - i r)^4} + \frac{i}{4 (a c + i r)^4} \,.
\end{multline}
Now the original integral $I_{l}$ has been converted to two integrals
of the form $\int P_{l}(c)/(ac\pm ir)^{n} dc$ where $n=4,5$. The power
$n$ may be reduced through integration by parts, i.e.~integrating
$(ac\pm ir)^{-n} dc$ while differentiating $P_{l}(c)$. After again
performing a partial fractions decomposition, this creates
two types of terms. First,
terms of the form $\int P_{l'}(c)/(ac\pm ir)^{n'} dc$ where
$n'=1,2,\ldots, n-1$
and $l' = l, l+1$, via~\cite{NIST:DLMF}
\begin{equation}
(1-x^{2})\frac{dP^{\mu}_{\nu}(x)}{dx}
=(\mu-\nu-1)P^{\mu}_{\nu+1}(x)+(\nu+1)xP^{\mu}_{\nu}(x)\,.
\end{equation}
Second, terms of the form
$\int P_{l'}(c)/(c\pm 1) dc$. The former terms with $n'=1$ may be
evaluated directly with Eq.~\eqref{eq:Pl-identity} and the other $n'$
may be repeatedly integrated by parts as just described. The remaining
terms of the form $\int P_{l'}(c)/(c\pm 1) dc$ can all be combined
together into integrals of the form
\begin{equation}
\int \frac{P_{l'-1}(c)-P_{l'+1}(c)}{1-c^2}dc\,.
\end{equation}
Here the integrand is subject to the identity
\begin{equation}
P_{n-1}(x)-P_{n+1}(x)=\frac{(2 n+1) \left(1-x^2\right) P_n'}{n(n+1)}
\end{equation}
which immediately yields
\begin{equation}
\int_{a}^{b} \frac{P_{l'-1}(c)-P_{l'+1}(c)}{1-c^2}dc = \frac{(2 n+1)}{n(n+1)}\left[P_n(b)-P_n(a)\right]\,.
\end{equation}
Applying these identities allows us to integrate $I_{l}(r)$ and gives
Eq.~\eqref{eq:I_l-using-Q}.

Though the argument of $Q_{l}$ is purely imaginary in
Eq.~\eqref{eq:I_l-using-Q}, the combinations $(Q_{l}(ix)+Q_{l}(-ix))$ and
$\frac{1}{i}(Q_{l}(+ix)-Q_{l}(-ix))$ are purely real. This can be seen
with the identity~\cite{NIST:DLMF}
\begin{equation}
Q_{l}(z) = \frac{\sqrt{\pi}\Gamma(l+1)}{(2z)^{l+1}\Gamma(l+\frac{3}{2})}
F\left(\frac{l+1}{2},\frac{l+2}{2};l+\frac{3}{2};\frac{1}{z^{2}}\right)\,.
\end{equation}
Using this identity gives Eq.~\eqref{eq:I_l-using-F} which is
manifestly real.

\section{Checks and comparisons}
\label{sec:checks-comparisons}
For any given $l$, typical computer algebra systems (such as
\textsc{Mathematica}) can perform the explicit integral $I_{l}(r)$,
since it is nothing but a rational polynomial function. We have
checked that Eq.~\eqref{eq:I_l-using-F} agrees with the explicit
evaluation of these integrals for a large number of $l$'s.

We have also compared our expressions (given in
Appendix~\ref{sec:source-moments-small-l}) with those given in KT. We
have verified the relationship
\begin{equation}
  S_{l}^{\text{KT}} = \frac{2l+1}{2} I_{l}
\end{equation}
where $S_{l}^{\text{KT}}$ are the expressions given in Appendix A of KT. This
suggests that KT have dropped the factor of $96 C$ (they scale all
dimensional quantities by $M$). Their expressions should be multiplied
by this factor, which they take as $3/2\pi$.
\newpage

\acknowledgments
The author would like to acknowledge Barry Wardell for helpful
discussions.
LCS acknowledges that support for this work was
provided by the National Aeronautics and Space Administration through
Einstein Postdoctoral Fellowship Award Number PF2-130101 issued by the
Chandra X-ray Observatory Center, which is operated by the Smithsonian
Astrophysical Observatory for and on behalf of the National
Aeronautics Space Administration under contract NAS8-03060.

\newpage
\appendix

\section{Source moments for small $l$}
\label{sec:source-moments-small-l}

\newcommand{\arctanar}{\arctan{\textstyle(\frac{a}{r})}}
\begin{align}
I_1={}& +2 (a^2 + r^2)^{-4} ra (r^{2}-a^2)\\
I_{3}={}&- {\textstyle \frac{1}{6}}(a^{2}+r^{2})^{-4} r a^{-3}\left[ 57
  a^6 + 73 a^4 r^2 \right.\\
&\qquad\left.{}+ 55 a^2 r^4 + 15 r^6\right] \nn\\
&+ {\textstyle\frac{5}{2}}a^{-4}\arctanar\nn\\
I_{5}={}&+{\textstyle\frac{1}{12}}(a^2 + r^2)^{-4}ra^{-5}
\left[
1047 a^8 + 5533 a^6 r^2 \right.\\
&\qquad\left. {}+ 9583 a^4 r^4 + 7035 a^2 r^6 + 1890 r^8
\right]\nn\\
 &-  {\textstyle\frac{35}{4}}a^{-6}(a^2 + 18 r^2)\arctanar\nn\\
I_{7}={}&- {\textstyle \frac{1}{16}}(a^2 + r^2)^{-4}ra^{-7}
\left[
4805 a^{10} + 56665 a^8 r^2 \right.\\
&\qquad\left. {}+188298 a^6 r^4 + 270354 a^4 r^6\right.\nn\\
&\qquad\left. {}+179025 a^2 r^8 + 45045 r^{10}
\right]\nonumber\\
& + {\textstyle\frac{315}{16}}a^{-8}(a^4 + 44 a^2 r^2 + 143 r^4)\arctanar\nn
\end{align}


\bibliographystyle{apsrev4-1}
\bibliography{CS-Kerr-scalar}

\end{document}